\documentclass[12pt]{article}
\usepackage{epsfig}
\usepackage{citesort}
\date{\today}
 \normalsize

\newcommand{\bmat}{\left(\begin{array}}
\newcommand{\emat}{\end{array}\right)}
\newcommand{\be}{\begin{equation}}
\newcommand{\ee}{\end{equation}}
\newcommand{\bea}{\begin{eqnarray}}
\newcommand{\eea}{\end{eqnarray}}

\def\NPB#1#2#3{Nucl. Phys. B{#1} (19#2) #3}

\def\Kah{{K$\ddot{a}$hler }}

\def    \be            {\begin{equation}}
\def    \ee            {\end{equation}}
\def    \bea           {\begin{eqnarray}}
\def    \eea           {\end{eqnarray}}

\def\ov{\overline}
\begin{document}
\renewcommand{\thefootnote}{\fnsymbol{footnote}}
\begin{titlepage}
\rightline{hep-th/0510196} \rightline{\today} \vspace{.5cm} {\large
\begin{center}
{\large \bf On Flux Compactification and Moduli Stabilization}
\end{center}}
\vspace{.3cm}
\begin{center}
A. Awad$^{1,2}$\footnote{aawad@ictp.it},
N.Chamoun$^{1,3}$\footnote{nchamoun@ictp.it}, and
S.Khalil$^{2,4}$\footnote{shaaban.khalil@guc.edu.eg}\\
\vspace{.3cm}

\emph{$^1$ The Abdus Salam ICTP, P.O. Box 586, 34100
Trieste, Italy.}\\
\emph{$^2$ Ain Shams University, Faculty of Science, Cairo 11566,
Egypt.}\\
\emph{$^3$ Physics Department, HIAST, P.O.Box 31983, Damascus,
Syria.}\\
\emph{$^4$ Department of Mathematics, German University in Cairo,
New
Cairo city, El Tagamoa El Khames, Egypt.}\\
\end{center}
\vspace{.5cm}
\begin{minipage}[h]{14.0cm}
\begin{center}
\small{\bf Abstract}\\[3mm]
\end{center}
%We analyze different approach to obtain de Sitter or Minkowiski
%non-SUSY vacua in type IIB string theory compactification with
%fluxes.
We study the effect of adding charged matter fields to both $D3$
and $D7$ branes in type IIB string theory compactification with
fluxes. Generically, charged matter fields induce additional terms
to the \Kah form, the superpotential and the D-terms. These terms
allow for minima with positive or zero cosmological constants,
even in the absence of non-perturbative effects. We show this
result first by decoupling the dilaton field along the lines of
the KKLT, and second by reincorporating it in the action with the
\Kah moduli.
\end{minipage}

\end{titlepage}
%%%%%%%%%%%%%%%%%%%%%%%%%%%%%%%%%%
\section{{\large \bf Introduction}}
In flux compactifications of type IIB supergravity, all the
complex structure moduli and the dilaton are generically fixed by
the non-trivial superpotential induced by the 3-form field
strengths\cite{Pol95}. However the \Kah moduli are not fixed by
the fluxes, and the resulting 4D model is of no-scale type.

Kachru, Kallosh, Linde and Trivedi (KKLT)'s approach \cite{kklt}
is the first explicit realization of 4D de Sitter space as a
solution to the low-energy equations of string theory where all
the moduli are stabilized. Their scenario consists of three main
stages. First, geometrical fluxes due to RR and NS-NS 3-form field
strengths are introduced to stabilize the dilaton $S$ and the
complex structure moduli $Z_i$ (CSM). Second, non-perturbative
effects due to gaugino condensation in the gauge theory on $D7$
branes \cite{Derendinger:1985kk,Dine:1985rz}, or $D3$ instantons
\cite{witsup} are used to stabilize the \Kah moduli. The resulting
potential has an AdS--SUSY minimum. In the final uplifting stage,
adding $\ov{D3}$ antibranes breaks SUSY explicitly and allows a
fine tuning of the cosmological constant to a small positive value
(de Sitter space).

Since the KKLT set up was proposed, it has been thoroughly
discussed and studied by many authors. It has been noticed
~\cite{Burgess:2003ic,alwis2} that the potential of the
anti-branes $\bar{D}3$
 should be added to the ten dimensional theory and not to the effective four dimensional action
 as was adopted in the KKLT set up.
 In this case, supersymmetry is explicitly broken and the resulting effective theory is not
 in a supergravity form, which leaves the theory uncontrollable.
 This problem was the motivation for many authors to consider alternative
 mechanisms in order to uplift the AdS vacuum to a Minkowiski or dS vacuum within the supergravity
 framework. In Ref.\cite{Burgess:2003ic}, fluxes of gauge fields
 that reside on D7 branes have been used to induce a positive
 D-term to the scalar potential allowing, thus, to obtain dS vacua.
 However, it has been emphasized in Ref. \cite{choi2} that one can
 not use D-terms to uplift the AdS SUSY vacuum to a dS non-SUSY one.

In this letter, we study the effect of charged matter and gauge
fluxes that live on D3 and/or D7 branes. The matter fields induce
additional terms to the \Kah function, the superpotential and the
D-terms. We show that, by incorporating these electric fluxes with
geometric fluxes in the presence of matter fields, one can stabilize
the complex structure moduli $Z_i$, the dilaton $S$, and the real
part of the \Kah modulus $T_R$, without invoking any
non-perturbative mechanism. We will not be concerned with the
imaginary part of $T$ which is analogous to the QCD $\theta$-term
and may be fixed by the non-perturbative axionic effect that breaks
Peccei-Quinn symmetry, see for example \cite{Ibanez:1991qh}. We also
find that the matter field contributions to the D-terms play a
crucial role in obtaining non-SUSY  Minkowiski or dS vacuua. We show
these results in the two most common scenarios: $i)$ The KKLT-like
case where $S$ and $Z_i$ are integrated out at a higher scale. $ii)$
The case in which both $S$ and $T$ are kept light in the effective
theory.

It is worth mentioning that in the later case $(ii)$ the KKLT set up
fails to obtain any local minimum to the potential~\cite{choi1}. In
contrast, we show that the new corrections due to matter
 fields and electric fluxes allow the existence of a local AdS non-SUSY
 minimum in both $(T,S)$ directions. In this analysis we assume that some of the matter
 fields acquire non-vanishing vevs which are higher than the moduli mass
 scale.

This letter is organized as follows. In section 2 we review
briefly the KKLT set up and its variations including the dilaton
in the effective action. Section 3 is devoted to analyzing the
impact of adding charged matter in $D3$ and/or $D7$ branes.  We
show that these matter fields modify the \Kah potential and induce
D-terms that lead to a non-SUSY Minkowski or de Sitter vacua, even
in the absence of non-perturbative effects and also regardless of
whether the dilaton is decoupled or not from the low energy
regime. Our conclusions are given in section 4.
%
%%%%%%%%%%%%%%%%%%%%%%%%%%%%%%%%%%%%%%%%%%%%%%%%%%%%%%%%%%%
%
\section{KKLT and its variants}
In this section we briefly review flux compactifications and the
KKLT approach and its variations for stablizing all moduli. In
type IIB theory, strings can have RR and NS-NS antisymmetric
3-form field strengths ($H_3$ and $F_3$ respectively) which can
wrap 3-cycles of the compactification manifold labeled by $P$ and
$Q$, leading to the following background fluxes

\bea
 \frac{1}{4\pi^2 \alpha'} \int_P F_3\ = L \, , \qquad
 \frac{1}{4\pi^2 \alpha'} \int_Q H_3\ = -K\,,
\eea
where $K$ and $L$ are integers. In the effective $4D$
supergravity, these geometric fluxes generate a superpotential for
the Calabi-Yau (CY) moduli, which is of the form \cite{GVW} \be W
~=~\int_{M} G_{3} \wedge \Omega \label{fluxsup} \ee where $\Omega$
is the holomorphic 3-form which depends on the CSM moduli $Z_i$
and $G_3 = F_3 - i S H_3$. The axion-dilaton field $S$ and the
overall scale modulus $T$ are defined in type IIB by:
\begin{eqnarray}
S=\frac{e^{-\phi}}{2\pi}+i c_0\, , \quad
T=\frac{e^{-\phi}}{2\pi}\left(M_{st}R\right)^4+i c_4\, ,
\end{eqnarray}
where $c_0$ and $c_4$ are the axions from the RR 0-form and
4-form, respectively, $e^{-\phi}=g_{st}$ denotes the string
coupling, $1/M_{st}^2$ is the string tension, and $R$ is the
compactification radius of the CY volume $V_{CY}\equiv (2\pi
R)^6$.

Giddings, Kachru and Polchinski (GKP) showed that these 3-form
fluxes can generically  stabilize the dilaton $S$ and all the CMS
moduli $Z_i$ \cite{GKP}. However, since the \Kah modulus $T$ does
not appear in the potential, it can not be fixed by the geometric
fluxes and the potential is of no-scale type. This partial fixing
of moduli in GKP framework can be understood by considering the
following tree-level \Kah potential: \bea & \label{e.dk} K = - 3
\log (T + \ov T) - \log (S + \ov S) -\log [-i\int_{M} \Omega
\wedge {\overline{\Omega}}]. \eea

The superpotential  (\ref{fluxsup}) and the \Kah potential
(\ref{e.dk}) lead to the following F-term potential: \bea V_F&=&
e^{K}\left( G^{I\bar{J}} D_{I}W \overline {D_J W} - 3|W|^2 \right)
=e^{K}(G^{i \bar{j}} D_i W \overline{D_j W}) \label{treepot} \eea
where $I$ and $J$ run over all moduli while $i$ and $j$ run over
dilaton and CSM moduli only. The covariant derivative is defined
as $D_I W =
\partial_I W + (\partial_I K) W$. Here $G^{I\bar{J}} = G^{-1}_{I\bar{J}}$
and $G_{I\bar{J}}$ is given by $G_{I\bar{J}} = K_{I\bar{J}} =
\partial_I \partial_{\bar{J}} K$. Note that $K^{T\bar{T}} \vert D_T
W \vert^2$ cancels with $3 \vert W \vert^2$, leaving the potential
$V_F$ independent of $T$. As can be seen from Eq.(\ref{treepot}),
the potential $V_F$ is positive definite, so that its global
minimum is at zero and hence the dilaton and CSM moduli are fixed
by the condition $D_i W=0$. This minimum is non-supersymmetric due
to the fact: $F_T \propto D_T W \propto W \neq 0$, {\it i.e.},
SUSY is broken by the $T$ field.

In order to fix $T$, KKLT considered a nonperturbative
superpotential, either generated by D3-brane instantons or by
gaugino condensation within a hidden non-abelian gauge sector on
the D7-branes. Since the dilaton and the CSM have been fixed at a
high scale, their contribution to the superpotential is a constant
$W_0$ and the total effective superpotential is given by \bea W
&=& W_0 + A e^{-a T},\eea where $A$ and $a$ are constants ($1/a$
is proportional to the beta function coefficient of the gauge
group in which the condensate occurs). The \Kah potential after
integrating out $S$ and $Z_i$ is reduced to \bea \label{KahlerNoS}
K&=&-3 \log (T+\ov{T}).\eea
 The new potential can now fix the field $T$ and one gets a
 supersymmetric AdS vacuum. To uplift this AdS minimum to a Minkowski or dS one, they added
 antibranes
 $\ov{D3}$.
 The $\ov{D3}$ effect amounts to an additional term in the scalar potential that is
 proportional to $\frac{1}{(T+\ov{T})^2}$, and for reasonable choices of
 parameters it yields de Sitter vacua.

As mentioned in the introduction, the $\ov{D3}$ explicitly breaks
supersymmetry and therefore the scalar potential is no longer in
its supergravity form. This complicates the analysis of the low
energy theory. A possible solution to overcome this problem has
been proposed in Ref.\cite{Burgess:2003ic} where the authors used
the $D$-term induced by the gauge fluxes on $D7$ branes. It turns
out that if the matter fields charged under the gauge group on
$D7$ branes are minimized at zero vevs, then the $D$-term gives
the same contribution to the scalar potential as $\ov{D3}$.
However, it is important to note that after including the
non-perturbative gaugino condensation in the second step of the
KKLT approach, supersymmetry is restored and the effective theory
is fully supersymmetric. Therefore, the potential $V_D$ is always
minimized at zero, as pointed out by several authors \cite{choi2},
and the $D$-term can not be used for uplifting the AdS SUSY vacua.

Another point of concern in the KKLT approach is the assumption
that the \Kah moduli are the only light moduli. As discussed in
\cite{choi1,alwis1}, there are
 situations where fluxes would keep the dilaton light
 while the CSM have string scale masses. In such cases,
 one can integrate out the CSM  but should leave the
 dilaton in the \Kah form and superpotential:
 \bea
\label{KahlerWithS} K &=& - 3 \log (T + \ov T) - \log (S + \ov S)
\eea \bea \label{WWithS} W &=& A + B S + C e^{-a T}\eea where
$A$,$B$, $C$ and $a$ are constants. In \cite{choi1} it was shown
that the AdS supersymmetric stationary point is in fact a saddle
point with instabilities along the moduli and axion directions,
and if one keeps $S$ as well as $T$ after including the gaugino
condensate then there are no local minima even in the presence of
a lifting potential of the form $\Delta V = {D \over (T + \ov
T)^{n_t}(S + \ov S)^{n_s}}$ where $n_t$ and $n_s$ are positive or
zero integers.
%
%%%%%%%%%%%%%%%%%%%%%%%%%%%%%%%%%%%%%%%%%%%%%%%%%%%%%%
%
\section{Charged matter fields and D-terms}
As advocated in the introduction, the problem of the no scale type
potential that is encountered in the KKLT setup can be overcome by
considering the effect of the charged chiral fields living on $D3$
and $D7$ branes in type IIB string. These fields generate a new
$T$-dependence in the \Kah potential which helps to stabilize the
$T$ moduli without assuming any non-perturbative mechanism.
Generically, in type IIB there are two types of massless $N=1$
chiral fields: closed string chiral fields (which include the
dilaton and moduli) and open string chiral fields which are
charged under the D-branes' gauge groups. In this setup we will
consider two types of branes, namely: $D3$ and $D7_i$ branes
(which are dual to $D9$ and $D5_i$ branes respectively). The index
$i = 1, 2, 3$, denotes the complex compact coordinate transverse
to the $D7$-brane world-volume. In this respect, we may have the
following charged matter fields: $C^3_i$ which arise from open
strings starting and ending on a $D3$ brane and $C^{7_i}_j$ which
come from open strings starting and ending on the same $7_i$. The
lowest order expansion of the \Kah potential in the matter field
is given by \cite{IMR} \be \label{KahlerWithSandmatter} K =
-\sum_{i=1}^{3} \log (T_i + \ov T_i) - \log (S + \ov S) +
\sum_{i=1}^3 \frac{\vert C^3_i \vert^2}{(T_i+\ov{T_i})}+
\sum_{i=1}^3 \frac{\vert C^{7_i}_i \vert^2}{(S+\ov{S})} +
\sum_{i,j,k=1}^3 d_{ijk} \frac{\vert C^{7_k}_j
\vert^2}{(T_i+\ov{T_i})}, \ee where $d_{ijk}=0$ if $i\neq j\neq
k$, otherwise $d_{ijk}=1$. One can also extract the renormalizable
contributions of the charged matter (i.e., the mass terms and
tri-linear couplings for the charged chiral superfields) to the
superpotential \cite{IMR}. Therefore, combining the charged matter
and geometric fluxes contributions leads to the following
superpotential \be W = A+BS+ g_3 C^3_1 C^3_2 C^3_3 +\sum_{i=1}^3
g_{7_i} C^{7_i}_1 C^{7_i}_2 C^{7_i}_3.\ee The Yukawa coupling
constants are given by the gauge couplings $g_3^2= 4\pi/\rm{Re} S$
and $g_{7}^2 = 4\pi/\rm{Re} T$. For simplicity, we assume that one
field, at most, on each type of $C^3_i$ and $C^{7_i}_i$ gets a
vev. Furthermore, the vevs of \Kah moduli acquire equal values and
the fields $C^{7_i}_j$ for $i\neq j$ are assumed to be zero. Here,
two comments are in order: i) The scale of the vevs of the charged
fields $C^{3,7}_i$ is assumed to be an intermediate, {\it i.e.},
below the scale of the CSM $Z_i$ and well above the scale of the
modulus $T$. ii) These vevs can be explicitly determined from the
full potential of the charged fields. It is interesting to find an
explicit example which leads to these desired vevs in type IIB,
however this beyond the scope of this letter and will be
considered elsewhere. In this case, the \Kah potential and the
superpotential take the form: \bea \label{KahlerWithSandmatter} K
&=& - 3 \log (T + \ov T) - \log (S
+ \ov S) + \frac{|\langle C_3\rangle|^2}{(T+\ov{T})}+ \frac{|\langle C_7 \rangle |^2}{(S+\ov{S})} \\
W &=& A+BS \eea

In addition to their contributions in the \Kah potential and the
superpotential, charged matter together with gauge fluxes give
rise to two different contributions to D-term. The first
contribution is coming from gauge fluxes which are known to induce
terms in the 4D $\cal N$ = 1 supersymmetric effective action
identified as Fayet-Iliopoulos (FI) D-terms \cite{Burgess:2003ic}.
Such terms have the following form in the $D7$ brane case \be
T_7\,\int_{\Gamma} F \wedge F = 2\,\pi {E^2 \over (T+\ov T)^3 }\ee
where $T_7$ is the $D7$ brane tension, $\Gamma$ is the 4-cycle
around which the $D7$ branes are wrapped and $E$ is the flux
strength. The other contribution is coming from matter fields.
Matter fields trigger spontaneous gauge symmetry breaking through
D-terms if they have non-vanishing vevs. The general form of
D-terms for the $D7$ and $D3$ branes can be expressed as \cite{u1}
\be \label{VDexpr}
 V_D =\, \frac{g_7^2}{2}\,\left( \sum_{i_7} q_{i_7} \Phi_{i_7}
 K_{i_7}+\xi_7\right)^2 + \frac{g_3^2}{2}\,\left( \sum_{i_3} q_{i_3} \Phi_{i_3}
 K_{i_3}+\xi_3\right)^2
\ee where $K_i$ is the derivative of the K\"ahler potential $K$
with respect to the matter fields $\Phi_i$ (a subset of $C^i$'s)
which has charge $q_i$ under the FI $U(1)$ group. The FI terms
$\xi_i$, where $i=3,7$ denotes the brane type, are given by
 $\xi_3=E_3 /\rm{Re} S$ and $\xi_7=E_7 /\rm{Re} T$~\cite{dsw}.

It is plausible to fine-tune the parameters: $\langle T \rangle$,
$\langle S \rangle$, $E$ and the vevs of the charged matter fields
($\langle \Phi_{i3,7} \rangle \equiv v_{3,7}$ ) so that gauge
symmetry is broken while SUSY remains exact. In Ref.
\cite{Burgess:2003ic}, it is assumed that charged matter fields
acquire a vanishing vev as a result of having only one type of
U(1) charge either positive or negative. Here, we will not assume
either of these situations, but, on the contrary, we will consider
the two D-term contributions mentioned above. Writing D-terms as a
function of the moduli fields, one gets the following expression
\bea \label{VD-term}
 V_D &=&\, \frac{1}{T+\ov{T}}\,\left( \frac{E_7}{T+\ov{T}}+\frac{F_7}{S+\ov{S}}
 \right)^2 + \frac{1}{S+\ov{S}}\,\left( \frac{E_3}{S+\ov{S}}+\frac{F_3}{T+\ov{T}}\right)^2
\eea where $E_7$($E_3$) is a measure of the strength of the flux
on D7 (D3) brane, and $F_7$ ($F_3$) denotes a `charged' vev, i.e.:
\bea F_{3,7}&\propto& {q_{3,7} |v_{3,7}|^2}. \eea If the matter
fields are minimized at $v_i = 0$, then the D-term, $V_D$, mimics
the effect of adding $\ov{D3}$ anti-branes, as pointed out in
Ref.\cite{Burgess:2003ic}. However, as we will show, having
non-vanishing vevs for these fields, {\it i.e.}, $F_i\neq 0$ and
also assuming the non-cancellation of the D-terms can play a
crucial role in obtaining a de Sitter vacuum.
%
%\subsection{Models with light Modulus $T$}
%

Now we present our results. In the generic case where the dilaton
field can be integrated out, we have the following scalar
potential \bea V&=& e^{K}\left( K^{T\ov T} D_{T}W \overline {D_T
W} - 3|W|^2 \right) + V_D \nonumber
\\ &=& e^K W_0
\frac{|v_3|^4}{\left(T+\ov{T}\right)\left[3\left(T+\ov{T}\right)+2|v_3|^2
\right]}+V_D
  \label{treepotDterm} \eea
 where \bea K&=& -3 \log(T+\ov{T})+ \frac{|v_3|^2}{T+\ov{T}}  +|v_7|^2\nonumber \\ W &=& W_0 \nonumber \\
 V_D &=& \frac{1}{T+\ov{T}}
 \left(\frac{E_7}{T+\ov{T}} + F_7\right)^2 +
 \left(E_3 + \frac{F_3}{T+\ov{T}}\right)^2\eea where the dilaton $ S$ was integrated out.

 As one can observe, the parameter space ($E_3$, $v_3$, $E_7$, and $v_7$) in this case is
 large, and for simplicity we choose that matter and
 gauge flux are present on the $D7$ branes but no matter are present on the $D3$ branes. Thus,
 the minimum of the above potential occurs
 at
 $t_{min}=-\frac{E_7}{2 F_7}$ with $E_7 . F_7 < 0$ (the other stationary point at $t=-\frac{3E_7}{2F_7}$
 would give a non-SUSY maximum). This is a non-SUSY
 Minkowski or de Sitter vacuum, according to whether or not we switch on gauge flux on the $D3$
 branes, since $V(t=t_{min})=E_3^2$. In this simple example
 the existence of non-vanishing charged vev $v_7$ is crucial in
order to obtain the above non-SUSY dS minimum. Here, the D-terms
alone (with both
 contributions from the FI term and the vev) is capable of stabilizing the volume modulus and producing a dS
vacuum without the need of non-perturbative effects.

In the case of non-vanishing $v_3$, one can show numerically the
existence of a non-SUSY dS minimum. Taking the values of the
parameters to be:
 \be \;\;\;F_3=F_7=0.1,\;\;\;E_7=0,\,\,E_3=-1.0,\;\;\;v_3=v_7=0.1,\;\;\;W_0=-0.01,\;\;\;\ee
 one can check that the minimum of the potential occurs at $t_{min}=0.05267$ and $V(t)|_{min}=0.0977$. The
 minimum is again a de Sitter vacuum.

 %For the other cases, the analytical expressions are complicated,
 %so we just checked numerically, for some chosen values of the parameters, if it was possible to get de Sitter
 %stable vacua. We summarize these results in Table \ref{table2}

%\begin{figure}
 %\epsfxsize=8.cm
  %\epsfysize=8.cm
  %\centerline{\epsfbox{fig4.eps}}
  %\caption{A non-supersymmetric stable de Sitter vacuum by the D-term. }
  %\label{fig2}
%\end{figure}

%\subsection{Models with light $S$ and $T$}

Now turning to the case where $T$ and $S$ are both left light. The
analytical expressions are difficult to obtain for the full
parameter space, but by setting $v_3=0$, and $E_3=0$, one can show
the existence of a non-SUSY Minkowiski vacua. For instance,
choosing \be \,\,\,v_3=E_3=0,\,\,\,q=-1,\,\,E_7=v_7^2,\,\,\,B={3 A
\over v_7^2 },\,\,\ee one finds a non-SUSY minimum at
$t_{min}=E_7,\;s_{min}=v_7^2$, which is indeed a Minkowiski
vacuum. It is important to notice that for the cases we have
discussed where SUSY is broken, one might expect a relevant
one-loop corrections to the potentials. Therefore one should not
rule out the possibility of having Minkowiski or AdS vacua since
these corrections can uplift them to dS vacua with small
cosmological constant as has been discussed in
Ref.\cite{Nilles:1983ge}.

%\bea A=-10,\;\;\;B=-1,\;\;\;
%&&F=-1,\;\;\;E=1,\;\;\;|<v>|^2=0.01\nonumber\eea For this choice
%of parameters we get a non-supersymmetric de Sitter vacuum at
%$t=0.5$, $s=10$.

\section{{\large \bf Conclusions}}
In this letter we have argued that it is possible to obtain de
Sitter or Minkowski non-SUSY vacua in type IIB string theory
compactification with fluxes without using neither
non-perturbative effects nor adding anti-branes $\bar{D}3$. We
have accomplished this by adding charged matter and gauge fluxes
on both $D7$ and $D3$ branes. We found that the matter field
contributions to the D-terms, the \Kah form and the superpotential
are crucial to stabilize the complex structure moduli $Z_i$, the
dilaton $S$, and the real part of the \Kah modulus $T_R$ with dS
or Minkowiski non-SUSY vacua. We showed that these results are
valid in scenarios where either $T$ or ($T$, $S$) are the only
light moduli in the low energy effective theory.

\section*{{\large \bf Acknowledgements}}
It is always a pleasure to thank B. Acharya, P. Argyres, K.Choi,
N. Mahajan and A. Shapere for enlightening discussions and useful
comments. Major part of this work was done within the Associate
Scheme of ICTP. S.K would like to thank CERN theory group for
their hospitality, where part of this work took place.

\end{document}